# Analytical Gradient Theory for Spin-Free State-Averaged Second-Order Driven Similarity Renormalization Group Perturbation Theory (SA-DSRG-MRPT2) and Its Applications for Conical Intersection Optimizations

Jae Woo Park[*]

*Department of Chemistry, Chungbuk National University (CBNU), Cheongju 28644, Korea*

**Abstract**

The second-order multireference driven similarity renormalization group perturbation theory (DSRG-MRPT2) theory provides an efficient means of correcting the dynamical correlation with the multiconfiguration reference function. The state-averaged DSRG-MRPT2 (SA-DSRG-MRPT2) method is the simplest means of treating the excited states with DSRG-MRPT2. In this method, the Hamiltonian dressed with dynamical correlation is diagonalized in the CASCI state subspace (SA-DSRG-MRPT2c) or the configuration subspace (SA-DSRG-MRPT2). This work develops the analytical gradient theory for spin-free SA-DSRG-MRPT2(c) with the density-fitting (DF) approximation. We check the accuracy of the analytical gradients against the numerical gradients. We present applications for optimizing minimum energy conical intersections (MECI) of ethylene and retinal model chromophores (PSB3 and RPSB6). We investigate the dependence of the optimized geometries and energies on the flow parameter and reference relaxations. The smoothness of the SA-DSRG-MRPT2(c) potential energy surfaces near the reference (CASSCF) MECI is comparable to the XMCQDPT2 one. These results render the SA-DSRG-MRPT2(c) theory a promising approach for studies of conical intersections.

---

[*] E-mail: jaewoopark@cbnu.ac.kr

## 1. INTRODUCTION

Multireference perturbation theories (MRPTs) provide efficient and accurate balanced descriptions of static and dynamical correlations. There are many variants of MRPT2. For example, complete active space perturbation theory (CASPT),[1-3] *N*-electron valence perturbation theory (NEVPT),[4-6] multireference Møller–Plesset theory (MRMP),[7-9] generalized van Vleck perturbation theory (GVVPT),[10-12] and static-dynamic-static perturbation theory (SDSPT)[13,14] have been presented over the last four decades.

The multireference-driven similarity renormalization group perturbation theory (DSRG-MRPT) can be considered another kind of MRPT.[15,16] This method is based on the perturbative analysis of the driven similarity renormalization group (DSRG)[15-24] energy and flow equations.[16] This method naturally alleviates the intruder state problem (which is present in CASPT and MRMP), and the internally contracted basis (which is required for internally contracted methods such as CASPT and NEVPT) is not required.

From a practical perspective, the DSRG-MRPT2 theory has other advantages over the other MRPT2 methods. First, all the DSRG-MRPT2 equations are solved noniteratively [this is the case for the CASPT2(D), NEVPT2, and MRMP2 methods]. Second, this theory requires up to only three-particle reduced density matrix elements (which is the case for MRMP2). Additionally, the three-cumulant [cu(3)] approximation, in which one neglects the three-particle terms, yields sufficient accuracies in many cases.[24] Moreover, one can include the relaxations of the reference function (CI coefficients) through dynamical correlation with the effective Hamiltonian constructed from the DSRG-MRPT2 calculations.[18]

In treating the electronically excited states, accurate descriptions of static and dynamical correlations are essential, particularly near a degeneracy. MRPTs are therefore widely used to study excited states.[25-27] (X)MCQDPT2,[28-31] (X)MS-CASPT2,[31-33] XDW-CASPT2,[34] and QD-NEVPT2[35] are among the widely used variants for studying excited states. For DSRG-MRPT2, Li, Evangelista, and coworkers developed the state-averaged DSRG-MRPT2 (SA-DSRG-MRPT2)[19], multistate DSRG-MRPT2 [(X)MS-DSRG-MRPT2], and its dynamically weighted (DW) variant (DW-DSRG-MRPT2).[36]

Many quantum chemical studies of electronically excited states involve geometry optimizations and molecular dynamics (MD) simulations.[26] Naturally, the corresponding applications greatly benefit from the availability of analytical nuclear derivatives.[37] Analytical gradient algorithms for (X)MCQDPT2,[38-41] (X)MS-CASPT2,[32,42-52] and QD-NEVPT2[50,53-55] were developed and applied for conical intersection searches and MD simulations. For DSRG theories, the single-reference DSRG-PT2 gradient[22] and the multireference single-state DSRG-MRPT2 gradient[24] were recently reported by Wang, Li, and Evangelista.[24] Therein, the authors suggested several future directions. Such directions included incorporations of the reference relaxations, applications of the density-fitting approximation for reducing the storage (memory) requirement, excited state extensions, and implementations of the higher-order MR-DSRG theories.[24]

This work incorporates the former three features into the DSRG-MRPT2 analytical first derivatives based on spin-free implementations.[23] Namely, we present the implementation of analytical gradient theory for spin-free state-specific and state-averaged DSRG-MRPT2 based on density fitting (DF). We implement analytical derivatives both with (SA-DSRG-MRPT2) and without (SA-DSRG-MRPT2c) so-called reference relaxation.[18] We apply the developed theory to optimize the minimum energy conical intersection (MECI) geometries of ethylene and retinal model chromophores (PSB3 and RPSB6). We discuss the dependence of optimized geometries and energies on the DSRG flow parameters. We also investigate the potential energy surfaces near the reference (CASSCF) surface crossing point to examine the reliability of SA-DSRG-MRPT2(c) theory for studies of the conical intersections. We distribute the resulting multithreaded computer codes as a patch on the open-source program package BAGEL.[56]

## 2. THEORY

In this section, we first briefly summarize the spin-free DSRG-MRPT2 theory, closely following the presentation by Li, Evangelista, and coworkers. Then, we present the analytical gradient theory for the single-state and state-averaged DSRG-MRPT2. For readability, we closely follow the notation in the DSRG literature.[15-22] For the orbitals, $p, q, r, s, \ldots, a, b, \ldots, i, j, \ldots, e, f, \ldots, u, v, \ldots, m, n, \ldots$ denote the general, particles (**P**), holes (**H**), virtual (**V**), active (**A**), and core (**C**) orbitals, respectively.

**DSRG-MRPT2.** Here, we provide a very brief review of the DSRG-MRPT2 theory. For more detailed derivations, one can refer to the previous literature by Li, Evangelista, and coworkers.[15-22] We assume that general reference function,

$$|\Phi\rangle = \sum_I c_I |I\rangle, \tag{1}$$

is used, where $|I\rangle$ is a single Slater determinant. The DSRG energy is

$$E(s) = \langle \Phi | \overline{H}(s) | \Phi \rangle, \tag{2}$$

where $s$ is the flow parameter and $\overline{H}(s)$ is the transformed Hamiltonian,

$$\overline{H}(s) = \hat{U}^\dagger(s) \hat{H} \hat{U}(s), \tag{3}$$

whose nondiagonal part is the source operator $\hat{R}(s)$,

$$\left[\overline{H}(s)\right]_N = \hat{R}(s). \tag{4}$$

The unitary transformation matrix $\hat{U}(s)$ is parameterized with the anti-Hermitian amplitude operator $\hat{A}(s)$ as

$$\hat{U}(s) = \exp[\hat{A}(s)]. \tag{5}$$

This operator is defined as

$$\hat{A}(s) = \sum_{k=1}^{n} \hat{A}_k(s) = \sum_{k=1}^{n} \hat{T}_k(s) - \hat{T}_k^{\dagger}(s), \tag{6}$$

where the $k$-body excitation operator is

$$\hat{T}_k(s) = \frac{1}{(k!)^2} \sum_{ij...}^{\mathbf{H}} \sum_{ab...}^{\mathbf{P}} t_{ab...}^{ij...}(s) \left\{ a_{ij...}^{ab...} \right\}. \tag{7}$$

Here, "{ }" denotes normal ordering of the operators. With perturbative analysis of the DSRG energy and flow equations (Eqs. 2 and 4),[15] one can derive the DSRG-MRPT2 theory.[16] The Hamiltonian is partitioned into

$$\hat{H} = \hat{H}^{(0)} + \xi \hat{H}^{(1)}, \tag{8}$$

where the zeroth-order Hamiltonian is defined as

$$\hat{H}^{(0)} = E_0 + \hat{F}^{(0)}. \tag{9}$$

This operator includes $E_0$, the reference energy, and

$$\hat{F}^{(0)} = \sum_m^{\mathbf{C}} \varepsilon_m \hat{E}_{mm} + \sum_u^{\mathbf{A}} \varepsilon_u \hat{E}_{uu} + \sum_e^{\mathbf{V}} \varepsilon_e \hat{E}_{ee}, \tag{10}$$

where the orbitals are semicanonicalized (the Fock matrix is diagonal in the core-core, active-active, and virtual-virtual subblocks). The Fock matrix elements are

$$f_{pq} = h_{pq} + \sum_{rs} \gamma_{rs} \left[ (pq|rs) - \frac{1}{2}(pr|qs) \right] = h_{pq} + [\mathbf{g}(\boldsymbol{\gamma})]_{pq}, \tag{11}$$

and $\varepsilon_p$ is the $p$-th eigenvalue of the Fock matrix. The first-order Hamiltonian is

$$\hat{H}^{(1)} = \hat{F}^{(1)} + \hat{V}^{(1)}. \tag{12}$$

The one-body operator $\hat{F}^{(1)}$ and the two-body operator $\hat{V}^{(1)}$ include the off-diagonal contributions of the Fock operator and all two-body interactions, respectively. Then, the energy and flow equations with the dependence on the perturbative parameter $\xi$ are

$$E(s,\xi) = \langle \Phi | \overline{H}(s,\xi) | \Phi \rangle \tag{13}$$

$$\hat{R}(s,\xi) = \left[ \overline{H}(s,\xi) \right]_N. \tag{14}$$

The source operator, the anti-Hermitian amplitude operator, and the energy are then expanded in the form of a power series. Collecting the terms in Eqs. 13 and 14 with the same power of $\xi$ gives expressions for the DSRG-MRPT amplitudes and energies. The spin-free amplitude equations are

$$T_{ijab} = V_{abij} \frac{1-\exp(-s\Delta_{ijab}^2)}{\Delta_{ijab}}, \tag{15}$$

$$t_{ia} = \left[ f_{ia} + \frac{1}{2} \sum_{uvw} \gamma_{uv} \left( U_{ivaw} f_{wu} - U_{iwau} f_{vw} \right) \right] \frac{1-\exp(-s\Delta_{ia}^2)}{\Delta_{ia}} \equiv f_{ia}^{\text{eff}} \frac{1-\exp(-s\Delta_{ia}^2)}{\Delta_{ia}}, \tag{16}$$

where $V_{abij} = (ai|bj)$, the denominators are $\Delta_{ijab} = \varepsilon_i + \varepsilon_j - \varepsilon_a - \varepsilon_b$ and $\Delta_{ia} = \varepsilon_i - \varepsilon_a$, $\gamma_{uv} = \langle \Phi | \hat{E}_{uv} | \Phi \rangle$, and we define the antisymmetrized two-electron amplitude **U** as

$$U_{ijab} = 2T_{ijab} - T_{ijba}. \tag{17}$$

Then, the second-order correction to the energy is

$$\begin{aligned} E^{(2)}(s) &= \langle \Phi | \left[ \hat{H}^{(1)}, \hat{A}^{(1)}(s) \right] | \Phi \rangle + \frac{1}{2} \langle \Phi | \left[ \left[ \hat{H}^{(0)}, \hat{A}^{(1)}(s) \right], \hat{A}^{(1)}(s) \right] | \Phi \rangle \\ &= \frac{1}{2} \langle \Phi | \left[ \tilde{H}^{(1)}, \hat{A}^{(1)}(s) \right] | \Phi \rangle \end{aligned}, \tag{18}$$

where the modified first-order effective Hamiltonian, $\tilde{H}^{(1)}(s) = \hat{H}^{(1)}(s) + \hat{R}^{(1)}(s)$, is defined with the modified integrals

$$\tilde{v}_{abij} = V_{abij}\left[1 + \exp\left(-s\Delta_{ijab}^2\right)\right], \tag{19}$$

$$\tilde{f}_{ai} = f_{ai} + t_{ia}\exp\left(-s\Delta_{ia}^2\right). \tag{20}$$

**Analytical Gradient Theory: Single-State DSRG-MRPT2.** The first implementation of the DSRG-MRPT2 analytical gradient was recently presented by Wang, Li, and Evangelista.[24] For completeness, let us rederive the expressions for the analytical first-order derivative of the spin-free DSRG-MRPT2 energy. As in the cases of other MRPT2 theories,[12,32,37-55] the analytical gradient can be elegantly derived using the Lagrangian formalism. Including all the quantities in the DSRG-MRPT2 energy, the single-state DSRG-MRPT2 Lagrangian, which is stationary with respect to all parameters in the DSRG-MRPT2 energy, is

$$\begin{aligned}
L_{\mathrm{DSRG}} &= E^{(0)} + E^{(2)} \\
&+ \Lambda_{ai}^f\left[f_{ai} + f_{ia}^{\mathrm{eff}}\exp(-s\Delta_{ia}^2) - \tilde{f}_{ai}\right] + \Lambda_{ia}^t\left[f_{ia}^{\mathrm{eff}}\frac{1-\exp(-s\Delta_{ia}^2)}{\Delta_{ia}} - t_{ia}\right] \\
&+ \Lambda_{ia}^F\left[(f_{ia} + \frac{1}{2}\gamma_{uv}(U_{ivaw}f_{wu} - U_{iwau}f_{vw})) - f_{ia}^{\mathrm{eff}}\right] + \Lambda_{ijab}^U\left[2T_{ijab} - T_{ijba} - U_{ijab}\right] \\
&+ \Lambda_{ijab}^T\left[V_{abij}\frac{1-\exp(-s\Delta_{ijab}^2)}{\Delta_{ijab}} - T_{ijab}\right] + \Lambda_{abij}^v\left[V_{abij}(1+\exp(-s\Delta_{ijab}^2)) - \tilde{v}_{abij}\right], \\
&+ \Lambda_{ijab}^\Delta\left[\varepsilon_i + \varepsilon_j - \varepsilon_a - \varepsilon_b - \Delta_{ijab}\right] + \Lambda_{ia}^\Delta\left[\varepsilon_i - \varepsilon_a - \Delta_{ia}\right] \\
&+ \Lambda_{uvw,xyz}^{(3)}\left[\langle\Phi|\hat{E}_{uvw,xyz}|\Phi\rangle - \Gamma_{uvw,xyz}^{(3)}\right] + \Lambda_{uv,xy}^{(2)}\left[\langle\Phi|\hat{E}_{uv,xy}|\Phi\rangle - \Gamma_{uv,xy}^{(2)}\right] \\
&+ \Lambda_{uv}^{(1)}\left[\langle\Phi|\hat{E}_{uv}|\Phi\rangle - \gamma_{uv}\right]
\end{aligned} \tag{21}$$

where the Einstein summation convention is again employed. All the multipliers, denoted with $\Lambda$, are obtained sequentially by setting the Lagrangian to be stationary, as none of these quantities are coupled. For example, $\mathbf{\Lambda}^f$ and $\mathbf{\Lambda}^t$ are obtained by solving

$$0 = \frac{\partial}{\partial \tilde{f}_{ai}}\left\{E^{(0)} + E^{(2)} + \Lambda_{ai}^f\left[f_{ai} + f_{ia}^{\mathrm{eff}}\exp(-s\Delta_{ia}^2) - \tilde{f}_{ai}\right]\right\}, \tag{22}$$

$$0 = \frac{\partial}{\partial t_{ia}} \left\{ E^{(0)} + E^{(2)} + \Lambda^t_{ia} \left[ f^{\text{eff}}_{ia} \frac{1-\exp(-s\Delta^2_{ia})}{\Delta_{ia}} - t_{ia} \right] \right\}, \quad (23)$$

which results in

$$\Lambda^f_{em} = 2t_{me}, \Lambda^f_{av} = \sum_w t_{wa}\gamma_{vw}, \Lambda^f_{wm} = \sum_v t_{mv}\eta_{wv}, \Lambda^f_{ex} = \sum_{uvy} T_{uvey}\lambda_{xyuv}, \Lambda^f_{vm} = -\sum_{uxy} T_{umxy}\lambda_{xyuv} \quad (24)$$

$$\Lambda^t_{me} = 2\tilde{f}_{em}, \Lambda^t_{wa} = \sum_v \tilde{f}_{av}\gamma_{vw}, \Lambda^t_{mv} = \sum_w \tilde{f}_{wm}\eta_{vw}, \Lambda^t_{ue} = \sum_{xvy} \tilde{v}_{exvy}\lambda_{xyuv}, \Lambda^t_{mx} = -\sum_{uvy} \tilde{v}_{uvmy}\lambda_{xyuv}. \quad (25)$$

Then, the next multiplier $\Lambda^F$ is calculated by solving

$$\begin{aligned}
0 = \frac{\partial}{\partial f^{\text{eff}}_{ia}} &\left\{ E^{(0)} + E^{(2)} + \Lambda^f_{ai}\left[ f_{ai} + f^{\text{eff}}_{ia}\exp(-s\Delta^2_{ia}) - \tilde{f}_{ai} \right] \right.\\
&\left. +\Lambda^t_{ia}\left[ f^{\text{eff}}_{ia}\frac{1-\exp(-s\Delta^2_{ia})}{\Delta_{ia}} - t_{ia} \right] + \Lambda^F_{ia}\left[ (f_{ia} + \frac{1}{2}\gamma_{uv}(U_{ivaw}f_{wu} - U_{iwau}f_{vw})) - f^{\text{eff}}_{ia} \right] \right\},
\end{aligned} \quad (26)$$

and the solution is

$$\Lambda^F_{ia} = \Lambda^f_{ai}\exp(-s\Delta^2_{ia}) + \Lambda^t_{ia}\frac{1-\exp(-s\Delta^2_{ia})}{\Delta_{ia}}. \quad (27)$$

The other multipliers [$\Lambda^U, \Lambda^T, \Lambda^v, \Lambda^{(3)}, \Lambda^{(2)}$, and $\Lambda^{(1)}$] are computed in a similar way, basically by collecting all the terms that are multiplied with the variables $\mathbf{U}$, $\mathbf{T}$, $\tilde{\mathbf{v}}$, $\Gamma^{(3)}$, $\Gamma^{(2)}$, and $\gamma$. The exceptions are the multipliers $\Lambda^\Delta$, which take a form

$$\begin{aligned}
\Lambda^\Delta_{ia} &= \Lambda^t_{ia} f^{\text{eff}}_{ia}\left[ 2s\exp(-s\Delta^2_{ia}) + \frac{\exp(-s\Delta^2_{ia})-1}{\Delta^2_{ia}} \right], \\
&\quad -2s\Delta_{ia}\Lambda^f_{ai}f^{\text{eff}}_{ia}\exp(-s\Delta^2_{ia})
\end{aligned} \quad (28)$$

$$\begin{aligned}
\Lambda^\Delta_{ijab} &= \Lambda^T_{ijab}V_{abij}\left[ 2s\exp(-s\Delta^2_{ijab}) + \frac{\exp(-s\Delta^2_{ijab})-1}{\Delta^2_{ijab}} \right], \\
&\quad -2s\Lambda^v_{abij}V_{abij}\Delta_{ijab}\exp(-s\Delta^2_{ijab})
\end{aligned} \quad (29)$$

due to the nonlinear dependence of the energy on the denominators $\Delta$. The final DSRG-MRPT2 Lagrangian can be rearranged as

$$L_{\text{DSRG}} = \sum_{pq} f_{pq} d^{(2)}_{pq} + \sum_{pqrs} V_{pqrs} D^{(2)}_{pqrs}$$
$$+ \sum_{xyzuvw} \Lambda^{(3)}_{xyz,uvw} \left[ \langle \Phi | \hat{E}_{xyz,uvw} | \Phi \rangle - \Gamma^{(3)}_{xyz,uvw} \right]$$
$$+ \sum_{uvxy} \Lambda^{(2)}_{uv,xy} \left[ \langle \Phi | \hat{E}_{uv,xy} | \Phi \rangle - \Gamma^{(2)}_{uv,xy} \right] \quad (30)$$
$$+ \sum_{uv} \Lambda^{(1)}_{uv} \left[ \langle \Phi | \hat{E}_{uv} | \Phi \rangle - \gamma_{uv} \right],$$

where

$$d^{(2)}_{ai} = \frac{1}{2} \left( \Lambda^{f}_{ai} + \Lambda^{F}_{ia} \right)$$
$$d^{(2)}_{ii} = \sum_{jab} \Lambda^{\Delta}_{ijab} + \Lambda^{\Delta}_{jiab} + \sum_{a} \Lambda^{\Delta}_{ia} \quad (31)$$
$$d^{(2)}_{aa} = -\left( \sum_{jab} \Lambda^{\Delta}_{ijab} + \Lambda^{\Delta}_{ijba} + \sum_{i} \Lambda^{\Delta}_{ia} \right),$$

$$D^{(2)}_{abij} = \Lambda^{T}_{ijab} \frac{1-\exp(-s\Delta^2_{ijab})}{\Delta_{ijab}} + \Lambda^{v}_{abij} \left[ 1 + \exp\left(-s\Delta^2_{ijab}\right) \right]. \quad (32)$$

The size of $\mathbf{D}^{(2)}$ could require a large amount of memory when the number of basis functions is more than 1000. We take advantage of the density-fitting approximation to reduce the memory requirement from $N^2_{\text{particle}} N^2_{\text{hole}}$ to $N_{\text{aux}} N_{\text{particle}} N_{\text{hole}}$,[17,41,44] where $N_{\text{aux}}$ is the number of auxiliary basis functions.

The total DSRG-MRPT2 Lagrangian is not stationary with respect to the variations in the orbital ($\mathbf{C}$) and CI coefficients ($\mathbf{c}$). The total Lagrangian, which is stationary with all the variations in $\mathbf{C}$ and $\mathbf{c}$, is

$$L = L_{\text{DSRG}} + \sum_{m \neq n} z^{c}_{mn} f_{mn} + \sum_{e \neq f} z^{c}_{ef} f_{ef}$$
$$+ \frac{1}{2} \sum_{pq} Z_{pq} \left[ \mathbf{A} - \mathbf{A}^{\dagger} \right]_{pq} + \sum_{u \neq v} Z^{c}_{uv} f_{uv} - \frac{1}{2} \sum_{pq} X_{pq} \left[ \left( \mathbf{C}^{\dagger} \mathbf{S} \mathbf{C} \right)_{pq} - \delta_{pq} \right], \quad (33)$$
$$+ \frac{1}{2} \sum_{I} z_{I} \langle I | \hat{H} - E^{(0)} | \Phi \rangle - x \left( \langle \Phi | \Phi \rangle - 1 \right),$$

where **A** defines the complete active space self-consistent field (CASSCF) convergence condition as

$$A_{tp} = 2\sum_{q}\left[h_{tq}\langle\Phi|\hat{E}_{pq}|\Phi\rangle + \sum_{rs}(tq|rs)\langle\Phi|\hat{E}_{pq,rs}|\Phi\rangle\right]. \tag{34}$$

Then, the so-called Z-vector equation takes a coupled form

$$\frac{\partial L}{\partial \kappa_{pq}} = 0, \tag{35}$$

$$\frac{\partial L}{\partial c_I} = 0, \tag{36}$$

where $\kappa_{rs}$ is the orbital rotation parameter that parameterizes the unitary orbital transform matrix as $\mathbf{V} = \exp(-\boldsymbol{\kappa})$. The source term for this equation is

$$Y_{pi} = \frac{\partial L_{\text{DSRG}}}{\partial \kappa_{pi}} = 2\left[\mathbf{f}\mathbf{d}^{(2)} + \mathbf{g}(\mathbf{d}^{(2)})\boldsymbol{\gamma} + \sum_{kl}\mathbf{D}^{kl}\mathbf{K}^{lk}\right]_{pi} + 2\sum_{jst}D_{st}^{ij}(ps|tj), \tag{37}$$

$$Y_{pa} = \frac{\partial L_{\text{DSRG}}}{\partial \kappa_{pa}} = 2\left[\mathbf{f}\mathbf{d}^{(2)} + \mathbf{g}(\mathbf{d}^{(2)})\boldsymbol{\gamma} + \sum_{kl}\mathbf{D}^{kl}\mathbf{K}^{lk}\right]_{pa}, \tag{38}$$

$$\begin{aligned}y_I = \frac{\partial L_{\text{DSRG}}}{\partial c_I} &= \sum_{xyzuvw}\Lambda^{(3)}_{uvw,xyz}\left[\langle I|\hat{E}_{uvw,xyz}|\Phi\rangle + \langle\Phi|\hat{E}_{uvw,xyz}|I\rangle\right] \\ &+ \sum_{xyuv}\Lambda^{(2)}_{uv,xy}\left[\langle I|\hat{E}_{uv,xy}|\Phi\rangle + \langle\Phi|\hat{E}_{uv,xy}|I\rangle\right] \\ &+ \sum_{uv}\Lambda^{(1)}_{uv}\left[\langle I|\hat{E}_{uv}|\Phi\rangle + \langle\Phi|\hat{E}_{uv}|I\rangle\right] + 2\sum_{uv}\langle I|\hat{E}_{uv}|\Phi\rangle\left[\mathbf{g}(\mathbf{d}^{(2)})\right]_{uv}\end{aligned}, \tag{39}$$

where $D_{rs}^{kl} = \frac{1}{2}D_{rkls}^{(2)}$ and $K_{rs}^{lk} = (rl|sk)$. We note that the derivative of the CASSCF energy is not included in these source terms, as this energy is already stationary with respect to the orbital and CI coefficients. In Eq. (37), the orbital gradient associated with the act-act rotations should be evaluated. This is because we compute the DSRG-MRPT2 energies and densities under the semicanonical condition of the active orbitals. The multipliers $\mathbf{z}^c$ is straightforwardly computed as

$$z_{mn}^c = -\frac{1}{2}\frac{Y_{mn} - Y_{nm}}{\varepsilon_m - \varepsilon_n}, \tag{40}$$

$$z_{ef}^c = -\frac{1}{2}\frac{Y_{ef} - Y_{fe}}{\varepsilon_e - \varepsilon_f}, \tag{41}$$

and these multipliers contribute to **Y** and **y** as

$$\mathbf{Y}' = \mathbf{Y} + 2\left[\mathbf{f}\mathbf{z}^c + \mathbf{g}(\mathbf{z}^{c)})\gamma\right], \tag{42}$$

$$y'_I = y_I + 2\sum_{uv}\left[\mathbf{g}(\mathbf{z}^c)\right]_{uv}\left\langle I\left|\hat{E}_{uv}\right|\Phi\right\rangle. \tag{43}$$

Then, one can solve the Z-vector equation with $\mathbf{Y}'$ and $\mathbf{y}'$ for **Z**, **z**, **X**, and $\mathbf{Z}^c$ using the method described in Ref. 40. With these evaluated multipliers, one can obtain the final gradients upon contraction with the integral derivatives. We also note that the core-core multiplier in Eq. 40 also accounts for the frozen core approximations.

**State-Averaged DSRG-MRPT2.** Next, let us give a brief synopsis of the state-averaged DSRG-MRPT2 (SA-DSRG-MRPT2) calculations. The state-averaged transformation of the Hamiltonian gives rise to the SA-DSRG-MRPT2 theory.[19] The basic equations are the same as the single-state case, but the state-averaged reduced density matrices (RDMs) and cumulants replace their state-specific counterparts.

To extract the energy of the electronic states, the DSRG-MRPT2 transformed Hamiltonian is diagonalized. The transformed Hamiltonian is

$$\overline{H}^{(1)}(s) = \hat{H}^{(1)} + \left[\hat{H}^{(0)}, \hat{A}^{(1)}(s)\right], \tag{44}$$

$$\overline{H}^{(2)}(s) = \left[\hat{H}^{(0)}, \hat{A}^{(2)}(s)\right] + \frac{1}{2}\left[\tilde{H}^{(1)}, \hat{A}^{(1)}(s)\right], \tag{45}$$

and the terms multiplied by the one- and two-particle density matrices are collected to form the one-electron and two-electron transformed integrals in the active space, $\bar{\mathbf{h}}$ and $\overline{(uv|xy)}$, respectivly. The effective core energy (scalar term) is[18]

$$E_{\text{core}} = E_{\text{DSRG-MRPT2}} - \bar{h}_{uv}\gamma_{uv}^{\text{SA}} \\ + \frac{1}{4}\left\{\left[2\overline{(vy|ux)} - \overline{(vy|xu)}\right]\gamma_{uv}^{\text{SA}}\gamma_{xy}^{\text{SA}} - 2\overline{(xy|uv)}\lambda_{uv,xy}^{(2),\text{SA}}\right\}, \quad (46)$$

while the state-averaged density matrices are evaluated as

$$\gamma_{uv}^{\text{SA}} = \sum_\alpha W_\alpha \left\langle \Phi_\alpha \left| \hat{E}_{uv} \right| \Phi_\alpha \right\rangle, \quad (47)$$

where $W_\alpha$ is the reference weight for the state $\alpha$. The state-averaged cumulants are evaluated according to Ref. 57 with the state-averaged density matrices. The weights are assumed to be identical for all the states in our current implementation. The elements $\bar{\mathbf{h}}$ and $\overline{(uv|xy)}$ are the Hamiltonian elements dressed with dynamical correlation.[3,21]

There are two schemes for obtaining the SA-DSRG-MRPT2 energies of each state. The first "contracted" scheme diagonalizes the Hamiltonian computed in the state space. Namely, diagonalizing the effective Hamiltonian

$$H_{\alpha\beta}^{(2)} = E_{\text{core}}\delta_{\alpha\beta} + \sum_{uv}\bar{h}'_{uv}\gamma_{uv}^{\alpha\beta} + \frac{1}{2}\sum_{uvxy}\overline{(uv|xy)}\Gamma_{uvxy}^{\alpha\beta}, \quad (48)$$

results in the energy,

$$E_P = \sum_{\alpha\beta} R_{\alpha P} H_{\alpha\beta}^{(2)} R_{\beta P}. \quad (49)$$

Note that the modified one-electron transformed integral[18]

$$\bar{h}'_{uv} = \bar{h}_{uv} - \frac{1}{2}\sum_{xy}\gamma_{xy}^{\text{SA}}[2\overline{(ux|vy)} - \overline{(ux|yv)}] \quad (50)$$

is used in this diagonalization. This scheme is abbreviated as SA-DSRG-MRPT2c. The second "uncontracted" scheme rediagonalizes the Hamiltonian in the determinantal or configuration state function (CSF) space. More explicitly, the Hamiltonian

$$H^{(2)}_{IJ} = E_{\text{core}}\delta_{IJ} + \sum_{uv} \bar{h}'_{uv} \langle I|\hat{E}_{uv}|J\rangle + \frac{1}{2}\sum_{uvxy} \overline{(uv|xy)} \langle I|\hat{E}_{uv,xy}|J\rangle, \qquad (51)$$

is diagonalized with conventional CASCI algorithms (e.g., Knowles–Handy[58]) so that

$$E_P = \sum_{IJ} c^{(2)}_{I,P} H^{(2)}_{IJ} c^{(2)}_{J,P}. \qquad (52)$$

The uncontracted calculation requires one more CASCI diagonalization. This scheme is the SA-DSRG-MRPT2 theory, which accounts for the relaxation of the reference[18] due to the dynamical correlation effect.

**Analytical Gradient Theory: SA-DSRG-MRPT.** Now, let us present the analytical gradient theory for the SA-DSRG-MRPT2 theory, both with and without the relaxed CI coefficients. The Lagrangian is

$$\begin{aligned}
L_{\text{DSRG},P} = E_P \\
+ \Lambda^h_{uv}\left[\bar{h}_{uv}(\tilde{\mathbf{f}},\mathbf{t},\mathbf{U},\mathbf{T},\tilde{\mathbf{v}}) - \bar{h}_{uv}\right] + \Lambda^g_{uv,xy}\left[\overline{(uv|xy)}(\tilde{\mathbf{f}},\mathbf{t},\mathbf{U},\mathbf{T},\tilde{\mathbf{v}}) - \overline{(uv|xy)}\right] \\
+ \Lambda^f_{ai}\left[f_{ai} + f^{\text{eff}}_{ia}\exp(-s\Delta^2_{ia}) - \tilde{f}_{ai}\right] + \Lambda^t_{ia}\left[f^{\text{eff}}_{ia}\frac{1-\exp(-s\Delta^2_{ia})}{\Delta_{ia}} - t_{ia}\right] \\
+ \Lambda^F_{ia}\left[(f_{ia} + \frac{1}{2}\gamma_{uv}(U_{ivaw}f_{wu} - U_{iwau}f_{vw})) - f^{\text{eff}}_{ia}\right] + \Lambda^U_{ijab}\left[2T_{ijab} - T_{ijba} - U_{ijab}\right] \\
+ \Lambda^T_{ijab}\left[V_{abij}\frac{1-\exp(-s\Delta^2_{ijab})}{\Delta_{ijab}} - T_{ijab}\right] + \Lambda^v_{abij}\left[V_{abij}(1+\exp(-s\Delta^2_{ijab})) - \tilde{v}_{abij}\right] \\
+ \Lambda^\Delta_{ijab}\left[\varepsilon_i + \varepsilon_j - \varepsilon_a - \varepsilon_b - \Delta_{ijab}\right] + \Lambda^\Delta_{ia}\left[\varepsilon_i - \varepsilon_a - \Delta_{ia}\right] \\
+ \Lambda^{(3)}_{uvw,xyz}\left[\sum_\alpha W_\alpha\langle\Phi_\alpha|\hat{E}_{uvw,xyz}|\Phi_\alpha\rangle - \Gamma^{\text{SA},(3)}_{uvw,xyz}\right] \\
+ \Lambda^{(2)}_{uv,xy}\left[\sum_\alpha W_\alpha\langle\Phi_\alpha|\hat{E}_{uv,xy}|\Phi_\alpha\rangle - \Gamma^{\text{SA},(2)}_{uv,xy}\right] + \Lambda^{(1)}_{uv}\left[\sum_\alpha W_\alpha\langle\Phi_\alpha|\hat{E}_{uv}|\Phi_\alpha\rangle - \gamma^{\text{SA}}_{uv}\right] \\
+ \Lambda^{(2),\alpha\beta}_{uv,xy}\left[\langle\Phi_\alpha|\hat{E}_{uv,xy}|\Phi_\beta\rangle - \Gamma^{\alpha\beta}_{uv,xy}\right] + \Lambda^{(1),\alpha\beta}_{uv}\left[\langle\Phi_\alpha|\hat{E}_{uv}|\Phi_\beta\rangle - \Gamma^{\alpha\beta}_{uv}\right] \qquad (53)
\end{aligned}$$

The multipliers are also sequentially obtained as in the case of the single-state DSRG-MRPT2

theory. For example, the multipliers $\Lambda^h$ and $\Lambda^g$ are evaluated as

$$\Lambda^h_{uv} = -\gamma^{SA}_{uv} + \sum_{\alpha\beta} R_{\alpha P} R_{\beta P} \gamma^{\alpha\beta}_{uv}, \tag{54}$$

$$\Lambda^g_{uv,xy} = \frac{1}{2}\gamma^{SA}_{xu}\gamma^{SA}_{yv} - \frac{1}{4}\gamma^{SA}_{yu}\gamma^{SA}_{xv} - \frac{1}{2}\lambda^{(2),SA}_{xyuv} + \frac{1}{2}\sum_{\alpha\beta} R_{\alpha P} R_{\beta P} \Gamma^{(2),\alpha\beta}_{uvxy}, \tag{55}$$

for SA-DSRG-MRPT2c, and

$$\Lambda^h_{uv} = -\gamma^{SA}_{uv} + \sum_{IJ} c^{(2)}_{I,P} c^{(2)}_{J,P} \left\langle I \left| \hat{E}_{uv} \right| J \right\rangle, \tag{56}$$

$$\Lambda^g_{uv,xy} = \frac{1}{2}\gamma^{SA}_{xu}\gamma^{SA}_{yv} - \frac{1}{4}\gamma^{SA}_{yu}\gamma^{SA}_{xv} - \frac{1}{2}\Gamma^{SA}_{xyuv} + \frac{1}{2}\sum_{IJ} c^{(2)}_{I,P} c^{(2)}_{J,P} \left\langle I \left| \hat{E}_{uv,xy} \right| J \right\rangle, \tag{57}$$

for SA-DSRG-MRPT2. Then, one computes the following multipliers using a similar method as in the single-state case. For example, the multipliers $\Lambda^f$ are evaluated as

$$\Lambda^f_{ai} = \frac{\partial E_P}{\partial \tilde{f}_{ai}} + \Lambda^h_{uv} \frac{\partial \overline{h}_{uv}}{\partial \tilde{f}_{ai}} + \Lambda^g_{uv,xy} \frac{\partial \overline{(uv|xy)}}{\partial \tilde{f}_{ai}}. \tag{58}$$

RDM-related multipliers [ $\Lambda^{(n),\alpha\beta}$ ] are also needed for the SA-DSRG-MRPT2c analytical gradients. These multipliers are

$$\Lambda^{(1),\alpha\beta}_{uv,xy} = R_{\alpha P} R_{\beta P} \overline{(uv|xy)}, \tag{59}$$

$$\Lambda^{(1),\alpha\beta}_{uv} = R_{\alpha P} R_{\beta P} \overline{h}_{uv}, \tag{60}$$

for SA-DSRG-MRPT2c. Then, the CI derivative of the SA-DSRG-MRPT2c energy is

$$y_{I,\alpha} = \frac{\partial L_{\text{DSRG}}}{\partial c_{I,\alpha}} = \sum_{uvwxyz} \Lambda^{(3)}_{uvw,xyz} \sum_{\alpha} W_\alpha \left[ \langle I | \hat{E}_{uvw,xyz} | \Phi_\alpha \rangle + \langle \Phi_\alpha | \hat{E}_{uvw,xyz} | I \rangle \right]$$

$$+ \sum_{uvxy} \Lambda^{(2)}_{uv,xy} \sum_{\alpha} W_\alpha \left[ \langle I | \hat{E}_{uv,xy} | \Phi_\alpha \rangle + \langle \Phi_\alpha | \hat{E}_{uv,xy} | I \rangle \right]$$

$$+ \sum_{uv} \Lambda^{(1)}_{uv} \sum_{\alpha} W_\alpha \left[ \langle I | \hat{E}_{uv} | \Phi_\alpha \rangle + \langle \Phi_\alpha | \hat{E}_{uv} | I \rangle \right] + 2 \sum_{\alpha} \sum_{uv} W_\alpha \langle I | \hat{E}_{uv} | \Phi_\alpha \rangle \left[ \mathbf{g}(\mathbf{d}^{(2)}) \right]_{uv}. \quad (61)$$

$$+ \sum_{uvxy} \Lambda^{(2),\alpha\beta}_{uv,xy} \langle I | \hat{E}_{uv,xy} | \Phi_\beta \rangle + \sum_{uvxy} \Lambda^{(2),\beta\alpha}_{uv,xy} \langle \Phi_\beta | \hat{E}_{uv,xy} | I \rangle$$

$$+ \sum_{uv} \Lambda^{(2),\alpha\beta}_{uv} \langle I | \hat{E}_{uv} | \Phi_\beta \rangle + \sum_{uv} \Lambda^{(2),\beta\alpha}_{uv} \langle \Phi_\beta | \hat{E}_{uv} | I \rangle$$

The SA-DSRG-MRPT2 energies are already optimal with respect to the CI coefficients, and their contributions to the CI derivative are zero. In other words, the computational cost for evaluating the last four terms in Eq. 61 is saved for SA-DSRG-MRPT2. The total Lagrangian is modified to reflect the state averaging as

$$L = L_{\text{DSRG}} + \sum_{m \neq n} z^c_{mn} f_{mn} + \sum_{e \neq f} z^c_{ef} f_{ef}$$

$$+ \frac{1}{2} \sum_{pq} Z_{pq} \left[ \mathbf{A} - \mathbf{A}^\dagger \right]_{pq} + \sum_{u \neq v} Z^c_{uv} f_{uv} - \frac{1}{2} \sum_{pq} X_{pq} \left[ (\mathbf{C}^\dagger \mathbf{S} \mathbf{C})_{pq} - \delta_{pq} \right] \quad (62)$$

$$+ \frac{1}{2} \sum_{\alpha} W_\alpha \left[ \sum_I z_{I,\alpha} \langle I | \hat{H} - E^{(0)} | \Phi_\alpha \rangle - x \left( \langle \Phi_\alpha | \Phi_\alpha \rangle - 1 \right) \right].$$

Then, the Z-vector equations are similarly solved as in the case of the state-specific DSRG-MRPT2 analytical gradients and the XMCQDPT2 analytical gradients.[40,41] Finally, we note that the so-called interstate couplings between the states $P$ and $Q$ can be simply obtained by substituting $E_P$ with the coupling elements ($\sum_{\alpha\beta} R_{\alpha P} H^{(2)}_{\alpha\beta} R_{\beta Q}$ and $\sum_{IJ} c^{(2)}_{I,P} H^{(2)}_{IJ} c^{(2)}_{J,Q}$ for SA-DSRG-MRPT2c and SA-DSRG-MRPT2, respectively) as described in Refs. 43 and 47. This quantity is useful for optimizing the conical intersection geometries.

## 3. NUMERICAL APPLICATIONS

Next, let us present the numerical applications of the SA-DSRG-MRPT2 analytical gradient theory. First, we test the accuracy of the current algorithm by comparing the analytical and finite-difference gradients. Then, we show the applications of the computer program for optimizing the MECIs of ethylene and retinal model chromophores (three double-bonded PSB3 and six double-bonded RPSB6). We also investigated the flow-parameter dependencies of the energies and optimized geometries of ethylene and PSB3. For comparative purposes, we also present the previous XMS-CASPT2, QD-NEVPT2,[55] and XMCQDPT2 results.[40] We evaluated the two-electron integrals using the DF approximation with the JKFIT basis sets[59] in all calculations. We used the frozen core approximations in SA-DSRG-MRPT2(c) calculations unless otherwise noted. We searched the MECIs with the gradient projection method established by Bearpark, Robb, and Schlegel.[60] We used the imaginary shift[61] of 0.2 $E_h$, and the intruder state avoidance (ISA) parameter[62] of 0.02 $E_h^2$ for XMS-CASPT2 and XMCQDPT2 computations, respectively, unless otherwise mentioned.

**Table 1.** Root-mean-squared differences between the SA-DSRG-MRPT2(c) analytical gradient and five-point finite difference gradient in a.u. ($E_h$ Bohr$^{-1}$). The DSRG flow parameter was set to 0.5 $E_h^{-2}$. The cc-pVTZ basis set was used. The frozen core approximation was not invoked in these tests. The test geometries are compiled in the Supporting Information.

|  | Active space | $N_{\text{state}}$ | State | SA-DSRG-MRPT2c | SA-DSRG-MRPT2 |
|---|---|---|---|---|---|
| LiF[a] | (6e, 4o) | 4 | $S_3$ | $2.47 \times 10^{-7}$ | $2.53 \times 10^{-7}$ |
| Ethylene[b] | (2e, 2o) | 3 | $S_1$ | $1.70 \times 10^{-7}$ | |
| Acrolein | (6e, 5o) | 1 | $S_0$ | $2.25 \times 10^{-6}$ | $3.42 \times 10^{-6}$ |
| Benzene | (6e, 6o) | 2 | $S_1$ | $1.92 \times 10^{-7}$ | $1.94 \times 10^{-7}$ |
| PSB3 | (6e, 6o) | 3 | $S_0$ | $4.34 \times 10^{-6}$ | $3.30 \times 10^{-6}$ |

[a] The def2-JKFIT basis set[63] was used for density-fitting approximation.
[b] The number of CSFs equals $N_{\text{state}}$, rendering the results of SA-DSRG-MRPT2(c) the same with and without reference relaxation.

**Accuracy of Analytical Gradient.** We tested the accuracy of SA-DSRG-MRPT2 analytical gradients for various molecules (lithium fluoride, ethylene, acrolein, benzene, and PSB3), active spaces, and electronic states with and without reference relaxation, and the results are shown in Table 1. The number of CSFs and $N_{\text{state}}$ are identical for ethylene, and the SA-DSRG-MRPT2 and SA-DSRG-MRPT2c results are the same. All the root-mean-squared errors between the analytical and five-point finite-difference gradients are smaller than $5.0 \times 10^{-6}$ a.u., which is similar to those from QD-NEVPT2[55] or XMCQDPT2.[40] The errors in PSB3 are about 20 times larger than in benzene. However, compared with the previous results, the magnitudes of errors in each system are similar to the XMCQDPT2 ones.[40] This similarity implies that these discrepancies are due to the quality of the CASSCF references. From these results, we conclude that the accuracy of our algorithm is satisfactory for optimizing the equilibrium and MECI geometries.

**Conical Intersection Optimizations of Ethylene.** The current algorithm is applied for the MECI optimizations of ethylene. We selected a small active space, ($2e$, $2o$), with three states for comparison to the previous results computed using multireference configuration interaction (MRCI),[64] XMS-CASPT2, QD-NEVPT2,[55] and XMCQDPT2.[40] As mentioned earlier, the SA-DSRG-MRPT2 and SA-DSRG-MRPT2c results are the same for this case. We used the aug-cc-pVTZ basis set. We optimized the ground state equilibrium geometry and four $S_0/S_1$ MECIs [Et, Et ($C_{3v}$), Hm, Pyr], as shown in Figure 1.

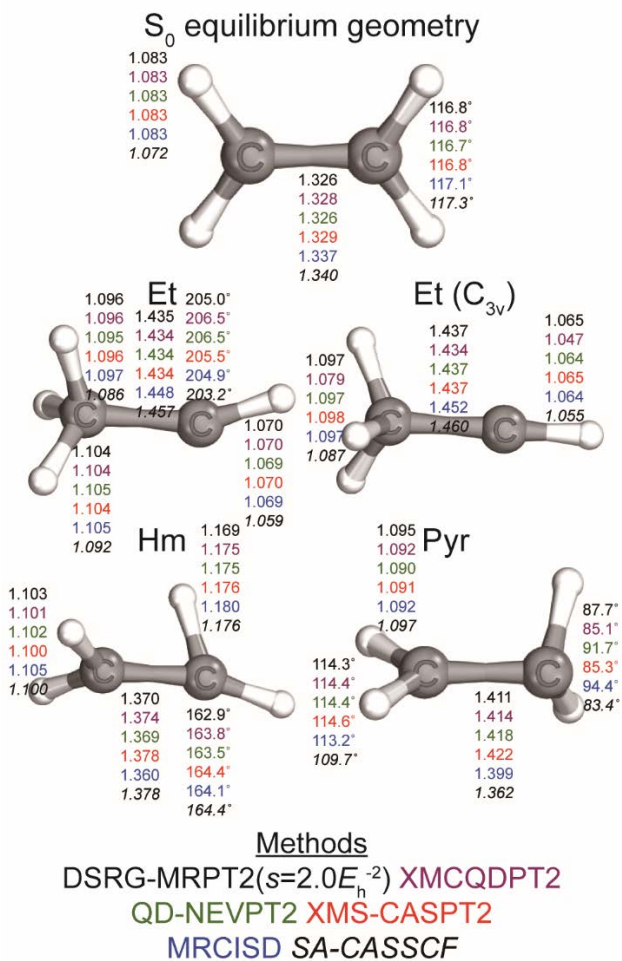

**Figure 1.** Geometries of the ethylene Franck–Condon point and MECIs optimized using SA-DSRG-MRPT2 with $s = 2.0\ E_h^{-2}$. The geometrical parameters (bond lengths in Å) for the other MRPT2 and MRCI methods are also displayed. The molecular graphics were generated with the program IboView.[65,66]

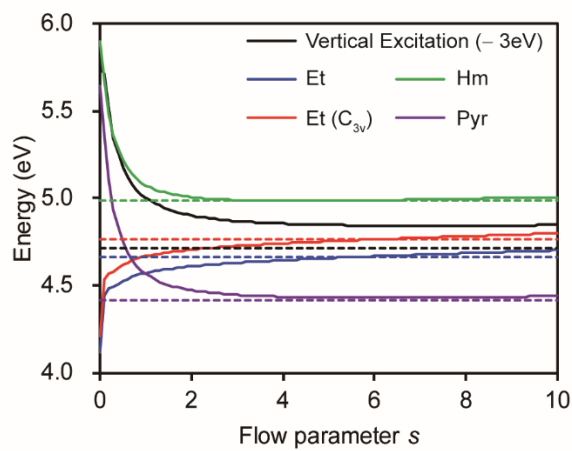

**Figure 2.** Dependence of the SA-DSRG-MRPT2 (solid) ethylene vertical excitation and conical intersection energies on the flow parameter. The $S_0$ energy at the Franck–Condon point is set to zero. The reference XMCQDPT2 values are also shown as dashed lines. For visibility, the vertical excitation energies are shifted by –3.0 eV.

**Table 2.** Energies (eV) at the equilibrium geometry and the optimized MECIs of ethylene for various methods. The $S_0$ energy at the Franck–Condon point is set to zero.

|  | DSRG flow parameter s ($E_h^{-2}$) | Vertical excitation energy | Et | Et ($C_{3v}$) | Hm | Pyr |
|---|---|---|---|---|---|---|
|  | 0.5 | 8.22 | 4.52 | 4.62 | 5.21 | 4.75 |
| SA-DSRG-MRPT2 | 1.0 | 8.05 | 4.57 | 4.66 | 5.07 | 4.57 |
|  | 2.0 | 7.95 | 4.61 | 4.70 | 5.01 | 4.47 |
| XMS-CASPT2[a] |  | 7.62 | 4.64 | 4.74 | 4.92 | 4.32 |
| QD-NEVPT2[55] |  | 7.64 | 4.74 | 4.86 | 5.13 | 4.51 |
| XMCQDPT2[40,b] |  | 7.71 | 4.66 | 4.76 | 4.99 | 4.42 |
| MRCISD+Q+RDP[64,c] |  | 7.79 | 4.56 | 4.69 | 5.20 | 4.46 |
| SA-CASSCF |  | 8.75 | 4.12 | 4.21 | 5.90 | 5.64 |

[a] The imaginary shift[49,61] of 0.2 $E_h$ was used.
[b] The ISA technique[62] was used with an ISA parameter of 0.02 $E_h^2$.
[c] The $S_0$ and $S_1$ energies can differ due to the state-specific +Q+RDP corrections. The $S_1$ energy is presented.

We should select the DSRG flow parameter $s$ carefully. The correlation energy is recovered to a greater extent when the higher the $s$ value is chosen. On the other hand, when the $s$ value is too high, the energy is more sensitive to the small energy denominators (intruder states). In other words, it is important to select an $s$ value that is high enough to catch large correlation energies but simultaneously low enough to avoid intruder states. (In the other MRPT2 methods, the shift[61,67] or ISA parameters[62] are used to avoid intruder states. The lowest parameter that avoids unphysical contributions of the small energy denominator is assumed to be the most accurate.) The first DSRG-MRPT2 study suggested that $s$ should be between [0.1, 1.0] $E_h^{-2}$,[16,21] and the first SA-DSRG-MRPT2 study employed $s = 0.5$ $E_h^{-2}$.[19] The recent implementation of a single-state DSRG-MRPT2 gradient used 1.0 $E_h^{-2}$.[24] We first checked the flow parameter value that is suitable for ethylene. The dependences of SA-DSRG-MRPT2 energies at the equilibrium and MECI geometries are displayed in Figure 2. The values at $s = 0.5$, 1.0, and 2.0 $E_h^{-2}$ are also shown in Table 2, along with the values computed with XMS-CASPT2, QD-NEVPT2, XMCQDPT2, and SA-CASSCF.

The SA-DSRG-MRPT2 vertical excitation energy is the highest among the MRPT2 and MRCI results. For $s=2.0$ $E_h^{-2}$, the vertical excitation energy is ~0.16 eV higher than the MRCISD+Q+RDP value,[64] and lower values of $s$ yield more significant errors. Nevertheless, the SA-DSRG-MRPT2 value has been lower than the SA-CASSCF value (8.75 eV) at any $s$ value we have tested. Therefore, we can conclude that the SA-DSRG-MRPT2 calculations recover the contributions of the dynamic correlation to the excitation energies to a reasonable degree at any $s$ values between [0.1, 2.0] $E_h^{-2}$. Using an $s$ value larger than ~10.0 $E_h^{-2}$ increases the computed vertical excitation energy values. This result implies that the recovered differential dynamical correlation effect becomes small with such a large $s$ value, so that qualitatively incorrect results are obtained. The energies at the MECIs are close to the previous MRPT2 values at s~2.0 $E_h^{-2}$. Overall, the SA-DSRG-MRPT2 energies are most comparable to the previous MRPT2 or MRCI results when $s$~2.0 $E_h^{-2}$.

We display the geometrical parameters of the FC point and MECIs at $s$=2.0 $E_h^{-2}$ in Figure 1. The SA-DSRG-MRPT2 optimized equilibrium geometry and MECIs share similar geometrical parameters with the other MRPT2 methods. The largest deviation in the bond length from the XMCQDPT2 value is ~0.006 Å in the Hm MECI, while many geometrical parameters agree within 0.005 Å and ~2 deg with the ones obtained with other PT2 methods. On the other hand, the SA-DSRG-MRPT2 geometries are significantly different from the SA-CASSCF geometries. In particular, the C-C bond lengths are either lengthened or shrunk by at least 0.005 Å by DSRG-MRPT2 corrections. Overall, our results for ethylene imply that the SA-DSRG-MRPT2 optimizations yield critical geometrical points that are comparable to the MRPT2 or MRCI results.

**Conical Intersection Optimizations of Retinal Model Chromophores: PSB3 and RPSB6.** Next, we applied the current algorithm to the retinal model chromophores, PSB3 and RPSB6. The active space included all the $\pi$ orbitals, resulting in (6$e$, 6$o$) and (12$e$, 12$o$) for these systems, respectively. We used the three lowest singlet states in the state-averaging scheme. We used the cc-pVTZ basis set. The ground state equilibrium geometry and two $S_0/S_1$ MECIs (11–12 and 13–14 twisted, abbreviated as 11–12 MECI and 13–14 MECI, respectively) shown in Figure 3 were optimized for both PSB3 and RPSB6.

As $N_{\text{state}} \neq N_{\text{CSF}}$, the SA-DSRG-MRPT2 and SA-DSRG-MRPT2c results differ for PSB3 and RPSB6. The vertical excitation energy at the FC point and the energies of the MECIs for both methods with varying flow parameters are shown in Figure 4. The values at $s$ = 0.5, 1.0, and 2.0 $E_\mathrm{h}^{-2}$ are also shown in Table 3, along with the values computed with XMS-CASPT2, QD-NEVPT2,[55] XMCQDPT2,[40] and SA-CASSCF. Again, in terms of energy, the SA-DSRG-MRPT2 values are closest to the XMCQDPT2 values at $s\sim2.0$ $E_\mathrm{h}^{-2}$. The vertical excitation energy and the 11–12 MECI energies without reference relaxation (SA-DSRG-MRPT2c) are similar to the values with reference relaxation. On the other hand, the 13–14 MECI energies are not close to the XMCQDPT2 values without reference relaxation, particularly for higher flow parameters. This result highlights the importance of reference relaxation for optimizing molecular geometries.

The geometrical parameters at the $S_0$ equilibrium geometry and MECIs are close to the MRPT2 parameters (Figure 3). To see how SA-DSRG-MRPT2(c) yields geometries that resemble the MRPT2 geometries, we also computed the Euclidean distances between the SA-CASSCF and XMCQDPT2 geometries and the SA-DSRG-MRPT2(c) geometries. The molecules were aligned to minimize the distances between them in the manner described in Ref. 68. The results for the $S_0$ equilibrium and 11–12 MECI based on $s$ variation are shown in Figure 5a. One can observe that SA-DSRG-MRPT2(c) resembles the XMCQDPT2 geometries more than the SA-CASSCF geometries for both the FC and 11–12 MECI. The SA-DSRG-MRPT2(c) geometries are the closest to the XMCQDPT2 geometries at $s\sim1.0$ $E_\mathrm{h}^{-2}$. When $s$ is higher than $\sim1.0$ $E_\mathrm{h}^{-2}$, the distance increases. The increase in the length is much smaller in SA-DSRG-MRPT2 than in SA-DSRG-MRPT2c. In other words, the reference relaxation improves the optimized geometries, mainly when $s$ is large. The trend in the distance contrasts with the result that the reference relaxation does not impact the energy at the 11–12 MECI relative to the $S_0$ energy (Figure 4). For the 13–14 MECI, the middle of the molecule is twisted, and the difference in the torsional angle along the 13–14 bond might exaggerate the Euclidean distance. Therefore, only one torsional parameter was compared (Figure 5b). As in the 11–12 MECI case, the SA-DSRG-MRPT2(c) torsional angle is similar to the XMCQDPT2 value when $s\sim1.0$ $E_h^{-2}$. Again, the SA-DSRG-MRPT2 value is closer to the XMCQDPT2 value than the SA-DSRG-MRPT2c value, especially when $s$ is large. Finally, when a large $s$ value ($s\sim3.0$ $E_h^{-2}$) is used, the SA-DSRG-MRPT2c optimizations converge to distinct geometrical points at varying $s$, and some bumps on the geometrical parameters appear. This implies that there are multiple minima with similar energies on the conical intersection seam.

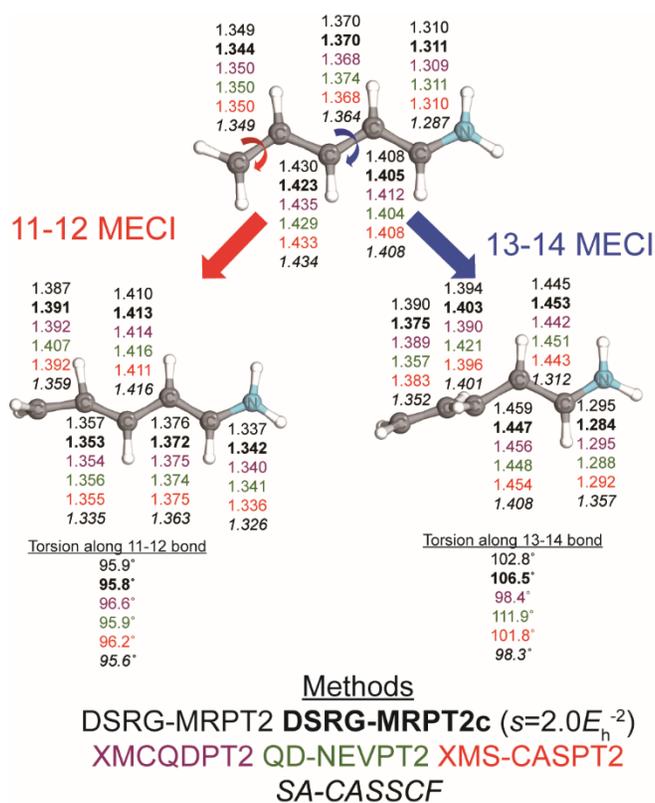

**Figure 3.** $S_0$ equilibrium geometry and MECIs of PSB3 optimized with SA-DSRG-MRPT2 using $s = 2.0$ $E_h^{-2}$.

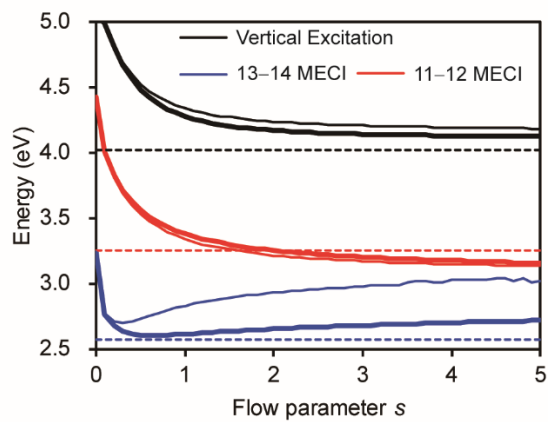

**Figure 4.** Dependence of SA-DSRG-MRPT2 (bold lines) and SA-DSRG-MRPT2c (thin solid lines) PSB3 vertical excitation and conical intersection energies on the flow parameter. The XMCQDPT2 values are also shown as dashed lines.

**Table 3.** Energies (eV) at the equilibrium geometry and optimized MECIs of PSB3 for various methods. The $S_0$ energy at the Franck–Condon point is set to zero.

|  | DSRG flow parameter $s$ ($E_h^{-2}$) | Vertical excitation energy | 11–12 MECI | 13–14 MECI |
|---|---|---|---|---|
|  | 0.5 | 4.48 | 3.56 | 2.60 |
| SA-DSRG-MRPT2 | 1.0 | 4.28 | 3.38 | 2.61 |
|  | 2.0 | 4.17 | 3.25 | 2.65 |
|  | 0.5 | 4.52 | 3.53 | 2.73 |
| SA-DSRG-MRPT2c | 1.0 | 4.34 | 3.34 | 2.83 |
|  | 2.0 | 4.24 | 3.22 | 2.93 |
| XMS-CASPT2[a] |  | 4.07 | 3.25 | 2.62 |
| QD-NEVPT2[55] |  | 4.24 | 3.08 | 3.15 |
| XMCQDPT2[40,b] |  | 4.02 | 3.26 | 2.57 |
| SA-CASSCF |  | 5.06 | 4.43 | 3.24 |

[a] The imaginary shift of 0.2 $E_h$ was used.
[b] The ISA technique was used with the ISA parameter of 0.02 $E_h^2$.

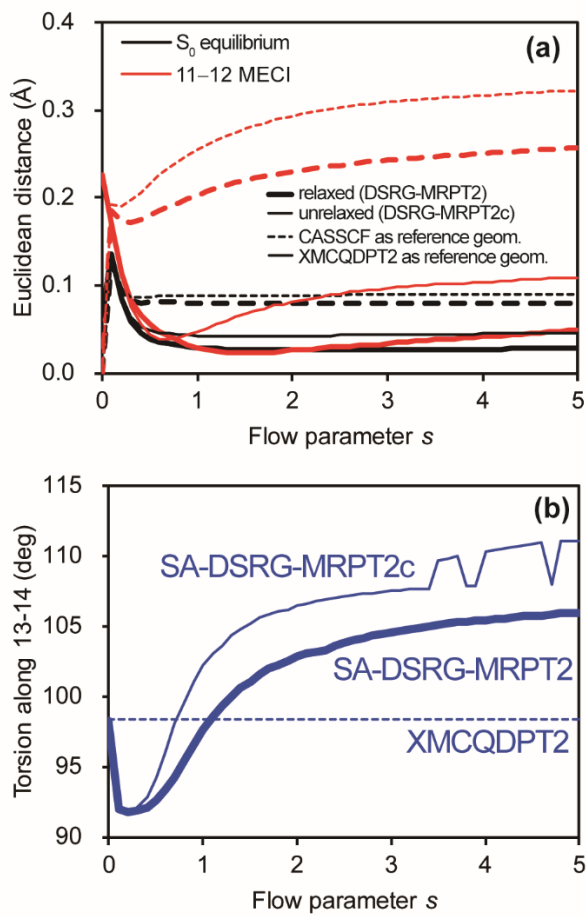

**Figure 5.** (a) Euclidean distances between the SA-DSRG-MRPT2(c) $S_0$ equilibrium (black) and 11–12 MECI (red) geometries and the SA-CASSCF (dashed) and XMCQDPT2 (solid) geometries. The presence of the reference relaxation is clarified through bold (SA-DSRG-MRPT2) and thin (SA-DSRG-MRPT2c) lines. (b) The torsional angle along the 13–14 bond at the 13–14 MECI geometries with (bold) and without (thin) reference relaxations. The XMCQDPT2 value is also shown as a dashed line.

Finally, let us demonstrate the MECI optimizations with SA-DSRG-MRPT2 for a larger system, RPSB6. The active space size is modest [(12$e$, 12$o$)], and the number of basis functions is 1108. We optimized the S$_0$ equilibrium geometry and MECIs using three $s$ values (0.5, 1.0, and 2.0 $E_\mathrm{h}^{-2}$). The resulting energies, compared with the previous XMCQDPT2 energies,[41] are presented in Table 4. The relative stabilities of the MECIs with respect to the S$_1$ energy at the Franck–Condon point at $s$~0.5 $E_\mathrm{h}^{-2}$ agrees well with the previous XMCQDPT2 results, while the vertical excitation energy is the most similar at $s$~2.0 $E_\mathrm{h}^{-2}$. From the relative stabilities of the MECIs, we assume that $s$=0.5 $E_\mathrm{h}^{-2}$ is the most suitable value for investigating the photodynamics of RPSB6. We present the geometrical parameters at these points in Figure 6. This demonstration shows that even though our implementation is not fully optimized, it is possible to use the SA-DSRG-MRPT2 analytical gradients with a density-fitting approximation to optimize the molecular geometries of the systems with $N_\mathrm{bas}$>1000 and a modest active space.

**Table 4.** Energies (eV) at the equilibrium geometry and optimized MECIs of RPSB6 for various methods. The S$_0$ energy at the Franck–Condon point is set to zero.

|  | DSRG flow parameter $s$ ($E_\mathrm{h}^{-2}$) | Vertical excitation energy | 11–12 MECI | 13–14 MECI |
|---|---|---|---|---|
| SA-DSRG-MRPT2 | 0.5 | 2.43 | 2.06 | 2.84 |
|  | 1.0 | 2.20 | 2.15 | 2.82 |
|  | 2.0 | 2.08 | 2.21 | 2.77 |
| XMCQDPT2[41][a] |  | 2.14 | 1.75 | 2.47 |
| SA-CASSCF |  | 3.36 | 2.39 | 2.21 |

[a] The ISA technique was used with the ISA parameter of 0.02 $E_\mathrm{h}^2$.

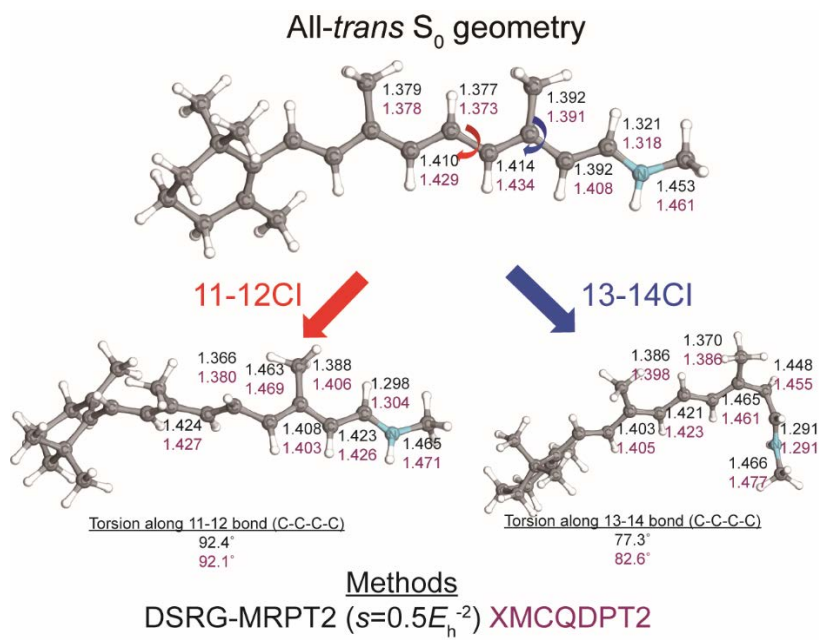

**Figure 6.** $S_0$ equilibrium geometry and MECIs of RPSB6 optimized with SA-DSRG-MRPT2 for $s = 0.5\ E_h^{-2}$.

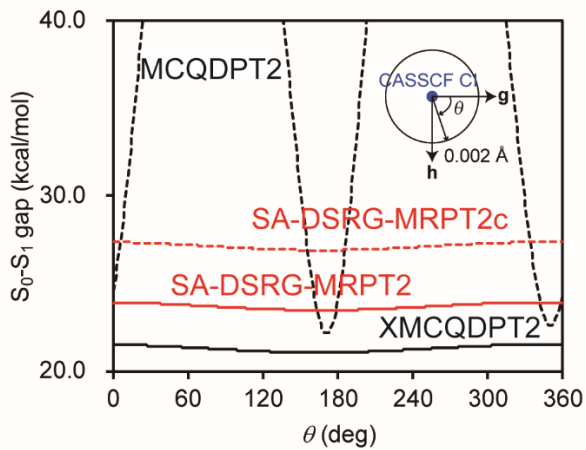

**Figure 7.** $S_0$ and $S_1$ energy gaps around the 0.002 Å loop centered at the CASSCF MECI, with MCQDPT2 (black dashed), XMCQDPT2 (black solid), SA-DSRG-MRPT2c (red dashed), and SA-DSRG-MRPT2 (red solid).

**Smoothness of SA-DSRG-MRPT2 Potential Energy Surfaces.** It is now widely acknowledged that the MCQDPT2, MS-CASPT2, and QD-NEVPT2 theories exhibit spurious behavior near the reference surface crossing points.[25,31,37,55,69-71] There are three sources of this behavior: (i) non-invariance of the zeroth-order Hamiltonian in the model space,[31,32] (ii) non-uniform zeroth-order Hamiltonian,[34,72] and (iii) non-uniform FOIS by electronic states.[31,55,69] The SA-DSRG-MRPT2(c) theory is based on the state-averaged DSRG transformation of the Hamiltonian, not relying on the multipartitioning scheme.[73,74] Therefore, one can expect that the SA-DSRG-MRPT2(c) potential energy surfaces (PESs) will not exhibit such unphysical behavior near the CASSCF conical intersections. Let us verify the smoothness of the PESs numerically.

The results of the PES scan along the loop encompassing the CASSCF 13–14 MECI of PSB3 with a radius of 0.002 Å is shown in Figure 7. The two vectors that lift the degeneracy near the CI [gradient difference vector (**g**) and interstate coupling vector (**h**)] define the loop. The amplitude of the gap energy oscillation is ~0.5 kcal/mol for all SA-DSRG-MRPT2, SA-DSRG-MRPT2c, and artifact-free XMCQDPT2 theories. This amplitude is also similar to the values obtained in the previous CASSCF, MRCISD, XMCQDPT2, or XMS-CASPT2 scans.[70,71] On the other hand, the amplitude of the gap energy oscillations obtained by the MCQDPT2 calculations is ~48 kcal/mol. This result strikingly contrasts the PESs obtained with the non-invariant (MCQDPT2) and invariant (XMCQDPT2) theories, and of course, the SA-DSRG-MRPT2(c) theories exhibit the behavior that is similar to the invariant perturbation theories. The contour of the $S_1$-$S_0$ energy gaps also shows the smooth PESs near the CASSCF MECI (Figure 8a). The potential energy surface is also smooth near the SA-DSRG-MRPT2(c) MECIs (Figure 8b). Notably, the differences between the $S_1$ energy computed with and without the cu(3) approximation are slight (below 0.1 kcal/mol), and this error is smooth as well (Figure 8c). Of course, the small error can be due to the small active space [(6$e$,6$o$)], and further investigations on the effects of cu(3) approximation are warranted. Overall, the SA-DSRG-MRPT2(c) theories yield smooth and physical PES near the reference surface crossing points. Therefore, given the small computational effort required, the SA-DSRG-MRPT2(c) methods can be a practical tool for investigating conical intersections.

## 4. CONCLUSION

In this work, we have derived the analytical gradient theory for spin-free SA-DSRG-MRPT2(c). With the density-fitting approximation, optimizations of the MECIs with more than 1,000 basis functions with modest active space [(12$e$, 12$o$)] were possible. This method has several practical advantages over the previous MRPT2 methods: The solution is noniteratively obtained, and the four-particle coupling terms are not required. Furthermore, reference relaxations allowed us to obtain better agreements with the other MRPT2 methods (XMS-CASPT2 and XMCQDPT2). The only additional computational demand for the reference relaxation is a single CASCI calculation. At the same time, the cost for the gradient is even slightly smaller due to the lack of the last four terms in Eq. 61. The SA-DSRG-MRPT2(c) theory did not exhibit any spurious behavior near the reference and/or PT2 conical intersection, making this approach promising to investigate the dynamics near the conical intersections. The source code for the SA-DSRG-MRPT2 analytical gradients is distributed as a patch on open-source BAGEL (`https://github.com/qsimulate-open/bagel`) version Jun 9 2021 at `http://sites.google.com/view/cbnuqbc/codes` under a GPL-v3 license.

## (a) Gap energies

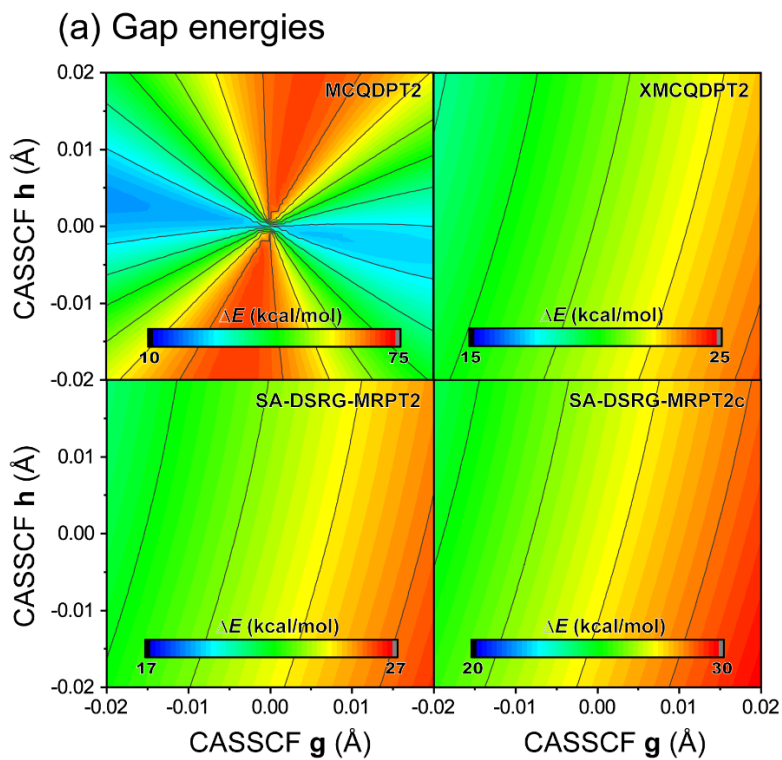

## (b) PES near the MECIs

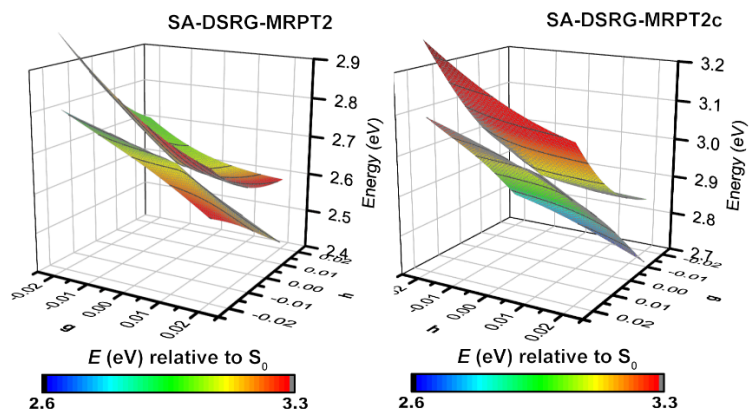

## (c) Effect of the cu(3) approximation

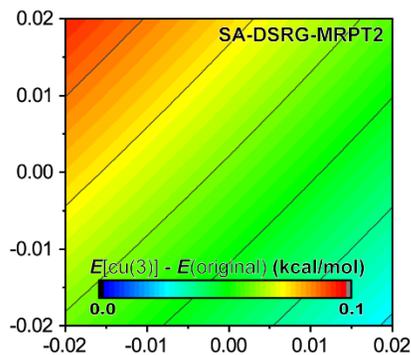

**Figure 8.** Smoothness of the PES near the CASSCF and SA-DSRG-MRPT2(c) MECIs. (a) The $S_0$-$S_1$ gap energies near the CASSCF MECIs obtained with (X)MCQDPT2 and SA-DSRG-MRPT2(c). Note the different color-scale between the MCQDPT2 contour and the others. (b) $S_0$ and $S_1$ PES near the SA-DSRG-MRPT2(c) MECIs. (c) The differences between the SA-DSRG-MRPT2 $S_1$ energy computed with and without the cu(3) approximation.

Our results showed a strong dependence of the energies and geometries on the flow parameters. To find an optimal value of the flow parameter *s,* more benchmarks are warranted for further photodynamic applications. Additionally, the analytical gradients for the higher-order DSRG(-MRPT) can be derived similarly or possibly using the automatic implementation technique (as suggested in Ref. 24). The current implementation is suitable only with the CASCI reference, and implementation using the approximated CI references[75] will also be needed to treat strongly correlated systems. Our implementation has yet to be fully optimized and parallelized. Improvement of the algorithm, particularly for massively parallel applications (MPI parallelization), is required. We are pursuing further investigations in these directions.

## ACKNOWLEDGMENTS

The (SA-)DSRG-MRPT2(c) energies were debugged using the existing implementation in FORTE[76] interfaced with the program package PSI4 1.4.[77] This work has been supported by the National Research Foundation (NRF) grant funded by the Korean government (MSIT) (Grant 2019R1C1C1003657) and by the POSCO Science Fellowship of POSCO TJ Park Foundation.

## REFERENCES

1.  Andersson, K.; Malmqvist, P.-Å.; Roos, B. O.; Sadlej, A. J.; Wolinski, K., Second-Order Perturbation Theory with a CASSCF Reference Function. *J. Phys. Chem.* **1990,** *94*, 5483-5488.

2.  Andersson, K.; Malmqvist, P.-Å.; Roos, B. O., Second-Order Perturbation Theory with a Complete Active Space Self-Consistent Field Reference Function. *J. Chem. Phys.* **1992,** *96*, 1218-1226.

3. Pulay, P., A Perspective on the CASPT2 Method. *Int. J. Quantum Chem.* **2011,** *111*, 3273-3279.

4. Angeli, C.; Cimiraglia, R.; Evangelisti, S.; Leininger, T.; Malrieu, J.-P., Introduction of N-Electron Valence States for Multireference Perturbation Theory. *J. Chem. Phys.* **2001,** *114*, 10252-10264.

5. Angeli, C.; Cimiraglia, R.; Malrieu, J.-P., N-electron Valence State Perturbation Theory: a Fast Implementation of the Strongly Contracted Variant. *Chem. Phys. Lett.* **2001,** *350*, 297-305.

6. Angeli, C.; Cimiraglia, R.; Malrieu, J.-P., N-Electron Valence State Perturbation Theory: A Spinless Formulation and an Efficient Implementation of the Strongly Contracted and of the Partially Contracted Variants. *J. Chem. Phys.* **2002,** *117*, 9138-9153.

7. Hirao, K., Multireference Møller-Plesset Method. *Chem. Phys. Lett.* **1992,** *190*, 374-380.

8. Hirao, K., Multireference Møller-Plesset Perturbation Theory for High-Spin Open-Shell Systems. *Chem. Phys. Lett.* **1992,** *196*, 397-403.

9. Hirao, K., State-Specific Multireference Møller-Plesset Perturbation Treatment for Singlet and Triplet Excited States, Ionized States and Electron Attached States of $H_2O$. *Chem. Phys. Lett.* **1993,** *201*, 59-66.

10. Khait, Y. G.; Song, J.; Hoffmann, M. R., Explication and Revision of Generalized Van Vleck Perturbation Theory for Molecular Electronic Structure. *J. Chem. Phys.* **2002,** *117*, 4133-4145.

11. Dudley, T. J.; Khait, Y. G.; Hoffmann, M. R., Molecular Gradients for the Second-Order Generalized Van Vleck Variant of Multireference Perturbation Theory. *J. Chem. Phys.* **2003,** *119*, 651-660.

12. Theis, D.; Khait, Y. G.; Hoffmann, M. R., GVVPT2 Energy Gradient Using a Lagrangian

Formulation. *J. Chem. Phys.* **2011**, *135*, 044117.

13. Lei, Y.; Liu, W.; Hoffmann, M. R., Further Development of SDSPT2 for Strongly Correlated Electrons. *Mol. Phys.* **2017**, *115*, 2696-2707.

14. Liu, W.; Hoffmann, M. R., SDS: The 'Static–Dynamic–Static' Framework for Strongly Correlated Electrons. *Theor. Chem. Acc.* **2014**, *133*, 1481-1492.

15. Evangelista, F. A., A Driven Similarity Renormalization Group Approach to Quantum Many-Body Problems. *J. Chem. Phys.* **2014**, *141*, 054109.

16. Li, C.; Evangelista, F. A., Multireference Driven Similarity Renormalization Group: A Second-Order Perturbative Analysis. *J. Chem. Theory Comput.* **2015**, *11*, 2097-2108.

17. Hannon, K. P.; Li, C.; Evangelista, F. A., An Integral-Factorized Implementation of the Driven Similarity Renormalization Group Second-Order Multireference Perturbation Theory. *J. Chem. Phys.* **2016,** *144*, 204111.

18. Li, C.; Evangelista, F. A., Driven Similarity Renormalization Group: Third-Order Multireference Perturbation Theory. *J. Chem. Phys.* **2017,** *146*, 124132.

19. Li, C.; Evangelista, F. A., Driven Similarity Renormalization Group for Excited States: A State-Averaged Perturbation Theory. *J. Chem. Phys.* **2018,** *148*, 124106.

20. Schriber, J. B.; Hannon, K. P.; Li, C.; Evangelista, F. A., A Combined Selected Configuration Interaction and Many-Body Treatment of Static and Dynamical Correlation in Oligoacenes. *J. Chem. Theory Comput.* **2018**, *14*, 6295-6305.

21. Li, C.; Evangelista, F. A., Multireference Theories of Electron Correlation Based on the Driven Similarity Renormalization Group. *Annu. Rev. Phys. Chem.* **2019**, *70*, 245-273.

22. Wang, S.; Li, C.; Evangelista, F. A., Analytic Gradients for the Single-Reference Driven Similarity Renormalization Group Second-Order Perturbation Theory. *J. Chem. Phys.* **2019,** *151*,

044118.

23. Li, C.; Evangelista, F. A., Spin-Free Formulation of the Multireference Driven Similarity Renormalization Group: A Benchmark Study of First-Row Diatomic Molecules and Spin-Crossover Energetics. *J. Chem. Phys.* **2021,** *155*, 114111.

24. Wang, S.; Li, C.; Evangelista, F. A., Analytic Energy Gradients for the Driven Similarity Renormalization Group Multireference Second-Order Perturbation Theory. *J. Chem. Theory Comput.* **2021,** *17*, 7666-7681.

25. Matsika, S., Electronic Structure Methods for the Description of Nonadiabatic Effects and Conical Intersections. *Chem. Rev.* **2021,** *121*, 9407-9449.

26. Lischka, H.; Nachtigallová, D.; Aquino, A. J. A.; Szalay, P. G.; Plasser, F.; Machado, F. B. C.; Barbatti, M., Multireference Approaches for Excited States of Molecules. *Chem. Rev.* **2018,** *118*, 7293-7361.

27. Szalay, P. G.; Müller, T.; Gidofalvi, G.; Lischka, H.; Shepard, R., Multiconfiguration Self-Consistent Field and Multireference Configuration Interaction Methods and Applications. *Chem. Rev.* **2012,** *112*, 108-181.

28. Nakano, H., Quasidegenerate perturbation theory with multiconfigurational self-consistent-field reference functions. *J. Chem. Phys.* **1993,** *99*, 7983-7992.

29. Nakano, H., MCSCF Reference Quasidegenerate Perturbation Theory with Epstein-- Nesbet Partitioning. *Chem. Phys. Lett.* **1993,** *207*, 372-378.

30. Granovsky, A. A. Table-Driven Implementation of the Multireference Perturbation Theories at Second Order. http://classic.chem.msu.su/gran/gamess/table_qdpt2.pdf (accessed Jun 8, 2021).

31. Granovsky, A. A., Extended Multi-Configuration Quasi-Degenerate Perturbation Theory:

The New Approach to Multi-State Multi-Reference Perturbation Theory. *J. Chem. Phys.* **2011,** *134*, 214113.

32. Shiozaki, T.; Győrffy, W.; Celani, P.; Werner, H.-J., Communication: Extended Multi-State Complete Active Space Second-Order Perturbation Theory: Energy and Nuclear Gradients. *J. Chem. Phys.* **2011,** *135*, 081106.

33. Finley, J.; Malmqvist, P. Å.; Roos, B. O.; Serrano-Andrés, L., The Multi-State CASPT2 Method. *Chem. Phys. Lett.* **1998,** *288*, 299-306.

34. Battaglia, S.; Lindh, R., Extended Dynamically Weighted CASPT2: the Best of Two Worlds. *J. Chem. Theory Comput.* **2020,** *16*, 1555-1567.

35. Angeli, C.; Borini, S.; Cestari, M.; Cimiraglia, R., A Quasidegenerate Formulation of the Second Order n-Electron Valence State Perturbation Theory Approach. *J. Chem. Phys.* **2004,** *121*, 4043-9.

36. Li, C.; Lindh, R.; Evangelista, F. A., Dynamically Weighted Multireference Perturbation Theory: Combining the Advantages of Multi-State and State-Averaged Methods. *J. Chem. Phys.* **2019,** *150*, 144107.

37. Park, J. W.; Al-Saadon, R.; MacLeod, M. K.; Shiozaki, T.; Vlaisavljevich, B., Multireference electron correlation methods: Journeys along potential energy surfaces. *Chem. Rev.* **2020,** *120*, 5878-5909.

38. Nakano, H.; Hirao, K.; Gordon, M. S., Analytic energy gradients for multiconfigurational self-consistent field second-order quasidegenerate perturbation theory (MC-QDPT). *J. Chem. Phys.* **1998,** *108*, 5660-5669.

39. Nakano, H.; Otsuka, N.; Hirao, K., Analytical Energy Gradients for Second-Order Multireference Perturbation Theory. In *Recent Advances in Computational Chemistry*, Hirao, K.,

Ed. World Scientific: Singapore, 1999; Vol. 4, pp 131-160.

40. Park, J. W., Analytical First-Order Derivatives of Second-Order Extended Multiconfiguration Quasi-Degenerate Perturbation Theory (XMCQDPT2): Implementation and Application. *J. Chem. Theory Comput.* **2020,** *16*, 5562-5571.

41. Park, J. W., Analytical Gradient Theory for Resolvent-Fitted Second-Order Extended Multiconfiguration Perturbation Theory (XMCQDPT2). *J. Chem. Theory Comput.* **2021,** *17*, 6122-6133.

42. Celani, P.; Werner, H.-J., Analytical Energy Gradients for Internally Contracted Second-Order Multireference Perturbation Theory. *J. Chem. Phys.* **2003,** *119*, 5044-5057.

43. Mori, T.; Kato, S., Dynamic electron correlation effect on conical intersections in photochemical ring-opening reaction of cyclohexadiene: MS-CASPT2 study. *Chem. Phys. Lett.* **2009,** *476*, 97-100.

44. Győrffy, W.; Shiozaki, T.; Knizia, G.; Werner, H.-J., Analytical Energy Gradients for Second-Order Multireference Perturbation Theory using Density Fitting. *J. Chem. Phys.* **2013,** *138*, 104104.

45. MacLeod, M. K.; Shiozaki, T., Communication: Automatic Code Generation Enables Nuclear Gradient Computations for Fully Internally Contracted Multireference Theory. *J. Chem. Phys.* **2015,** *142*, 051103.

46. Vlaisavljevich, B.; Shiozaki, T., Nuclear Energy Gradients for Internally Contracted Complete Active Space Second-Order Perturbation Theory: Multistate Extensions. *J. Chem. Theory Comput.* **2016,** *12*, 3781-3787.

47. Park, J. W.; Shiozaki, T., Analytical Derivative Coupling for Multistate CASPT2 Theory. *J. Chem. Theory Comput.* **2017,** *13*, 2561-2570.

48. Park, J. W.; Shiozaki, T., On-the-Fly CASPT2 Surface-Hopping Dynamics. *J. Chem. Theory Comput.* **2017,** *13*, 3676-3683.

49. Park, J. W.; Al-Saadon, R.; Strand, N. E.; Shiozaki, T., Imaginary Shift in CASPT2 Nuclear Gradient and Derivative Coupling Theory. *J. Chem. Theory Comput.* **2019,** *15*, 4088-4098.

50. Park, J. W., Analytical Gradient Theory for Strongly Contracted (SC) and Partially Contracted (PC) N-Electron Valence State Perturbation Theory (NEVPT2). *J. Chem. Theory Comput.* **2019,** *15*, 5417-5425.

51. Song, C.; Neaton, J. B.; Martínez, T. J., Reduced Scaling Formulation of CASPT2 Analytical Gradients Using the Supporting Subspace Method. *J. Chem. Phys.* **2021,** *154*, 014103.

52. Nishimoto, Y., Analytic Gradients for Restricted Active Space Second-Order Perturbation Theory (RASPT2). *J. Chem. Phys.* **2021,** *154*, 194103.

53. Nishimoto, Y., Analytic First-Order Derivatives of Partially Contracted N-Electron Valence State Second-Order Perturbation Theory (PC-NEVPT2). *J. Chem. Phys.* **2019,** *151*, 114103.

54. Nishimoto, Y., Locating Conical Intersections Using the Quasidegenerate Partially and Strongly Contracted NEVPT2 Methods. *Chem. Phys. Lett.* **2020,** *744*, 137219.

55. Park, J. W., Analytical Gradient Theory for Quasidegenerate N-Electron Valence State Perturbation Theory (QD-NEVPT2). *J. Chem. Theory Comput.* **2020,** *16*, 326-339.

56. Shiozaki, T., BAGEL: Brilliantly Advanced General Electronic-structure Library. *WIREs Comput. Mol. Sci.* **2018,** *8*, e1331.

57. Kutzelnigg, W.; Shamasundar, K. R.; Mukherjee, D., Spin-Free Formulation of Reduced Density Matrices, Density Cumulants and Generalised Normal Ordering. *Mol. Phys.* **2010,** *108*, 433-451.

58. Knowles, P. J.; Handy, N. C., A New Determinant-Based Full Configuration Interaction

Method. *Chem. Phys. Lett.* **1984,** *111*, 315-321.

59. Weigend, F., A Fully Direct RI-HF Algorithm: Implementation, Optimised Auxiliary Basis Sets, Demonstration of Accuracy and Efficiency. *Phys. Chem. Chem. Phys.* **2002,** *4*, 4285-4291.

60. Bearpark, M. J.; Robb, M. A.; Schlegel, H. B., A Direct Method for the Location of the Lowest Energy Point on a Potential Surface Crossing. *Chem. Phys. Lett.* **1994,** *223*, 269-274.

61. Forsberg, N.; Malmqvist, P. Å., Multiconfiguration Perturbation Theory with Imaginary Level Shift. *Chem. Phys. Lett.* **1997,** *274*, 196-204.

62. Witek, H. A.; Choe, Y. K.; Finley, J. P.; Hirao, K., Intruder State Avoidance Multireference Moller-Plesset Perturbation Theory. *J. Comput. Chem.* **2002,** *23*, 957-965.

63. Weigend, F., Hartree–Fock Exchange Fitting Basis Sets for H to Rn. *J. Comput. Chem.* **2008,** *29*, 167-175.

64. Barbatti, M.; Paier, J.; Lischka, H., Photochemistry of Ethylene: a Multireference Configuration Interaction Investigation of the Excited-State Energy Surfaces. *J. Chem. Phys.* **2004,** *121*, 11614-11624.

65. Knizia, G., Intrinsic Atomic Orbitals: An Unbiased Bridge between Quantum Theory and Chemical Concepts. *J. Chem. Theory Comput.* **2013,** *9*, 4834-4843.

66. Knizia, G.; Klein, J. E. M. N., Electron Flow in Reaction Mechanisms—Revealed from First Principles. *Angew. Chem. Int. Ed.* **2015,** *54*, 5518-5522.

67. Roos, B. O.; Andersson, K., Multiconfigurational Perturbation Theory With Level Shift - the $Cr_2$ Potential Revisited. *Chem. Phys. Lett.* **1995,** *245*, 215-223.

68. Rhee, Y. M., Construction of an Accurate Potential Energy Surface by Interpolation with Cartesian Weighting Coordinates. *J. Chem. Phys.* **2000,** *113*, 6021-6024.

69. Granovsky, A. A. On the Non-Invariance of QD-NEVPT2 Theory.

http://classic.chem.msu.su/gran/gamess/qdnevpt2-non-invariance.pdf (accessed Jan 13, 2022).

70. Gozem, S.; Huntress, M.; Schapiro, I.; Lindh, R.; Granovsky, A. A.; Angeli, C.; Olivucci, M., Dynamic Electron Correlation Effects on the Ground State Potential Energy Surface of a Retinal Chromophore Model. *J. Chem. Theory Comput.* **2012,** *8*, 4069-4080.

71. Gozem, S.; Melaccio, F.; Valentini, A.; Filatov, M.; Huix-Rotllant, M.; Ferre, N.; Frutos, L. M.; Angeli, C.; Krylov, A. I.; Granovsky, A. A.; Lindh, R.; Olivucci, M., Shape of Multireference, Equation-of-Motion Coupled-Cluster, and Density Functional Theory Potential Energy Surfaces at a Conical Intersection. *J. Chem. Theory Comput.* **2014,** *10*, 3074-3084.

72. Park, J. W., Single-State Single-Reference and Multistate Multireference Zeroth-Order Hamiltonians in MS-CASPT2 and Conical Intersections. *J. Chem. Theory Comput.* **2019,** *15*, 3960-3973.

73. Zaitsevskii, A.; Malrieu, J.-P., Multi-Partitioning Quasidegenerate Perturbation Theory. A New Approach to Multireference Møller-Plesset Perturbation Theory. *Chem. Phys. Lett.* **1995,** *233*, 597-604.

74. Malrieu, J.-P.; Heully, J.-L.; Zaitsevskii, A., Multiconfigurational Second-Order Perturbative Methods: Overview and Comparison of Basic Properties. *Theor. Chim. Act.* **1995,** *90*, 167-187.

75. Eriksen, J. J., The Shape of Full Configuration Interaction to Come. *J. Phys. Chem. Lett.* **2021,** *12*, 418-432.

76. Schriber, J. B.; Hannon, K.; Li, C.; Zhang, T.; Evangelista, F. A. *FORTE: A suite of quantum chemistry methods for strongly correlated electrons.* *https://github.com/evangelistalab/forte*, 2021.

77. Smith, D. G. A.; Burns, L. A.; Simmonett, A. C.; Parrish, R. M.; Schieber, M. C.;

Galvelis, R.; Kraus, P.; Kruse, H.; Di Remigio, R.; Alenaizan, A.; James, A. M.; Lehtola, S.; Misiewicz, J. P.; Scheurer, M.; Shaw, R. A.; Schriber, J. B.; Xie, Y.; Glick, Z. L.; Sirianni, D. A.; O'Brien, J. S.; Waldrop, J. M.; Kumar, A.; Hohenstein, E. G.; Pritchard, B. P.; Brooks, B. R.; Schaefer, H. F., III; Sokolov, A. Y.; Patkowski, K.; DePrince, A. E., III; Bozkaya, U.; King, R. A.; Evangelista, F. A.; Turney, J. M.; Crawford, T. D.; Sherrill, C. D., PSI4 1.4: Open-Source Software for High-Throughput Quantum Chemistry. *J. Chem. Phys.* **2020,** *152*, 184108.